\documentclass[pra,twocolumn,preprintnumbers,floatfix]{revtex4}
\usepackage{graphicx}
\usepackage{dcolumn}
\usepackage{bm}
\usepackage{latexsym,epsfig}
\usepackage{graphicx}
\usepackage{verbatim}
\usepackage{comment}
\usepackage{amsmath}
\usepackage{amssymb}
\usepackage{stmaryrd}
\usepackage{color}
\usepackage{epstopdf}
\usepackage{grffile}
\usepackage{lipsum}
\DeclareGraphicsExtensions{.eps}

\newcommand{\beq}{\begin{equation}}
\newcommand{\eeq}{\end{equation}}
\newcommand{\bea}{\begin{eqnarray}}
\newcommand{\eea}{\end{eqnarray}}
\newcommand{\ben}{\begin{eqnarray*}}
\newcommand{\een}{\end{eqnarray*}}
\newcommand{\bfig}{\begin{figure}}
\newcommand{\efig}{\end{figure}}

\usepackage{hyperref}
\hypersetup{
    colorlinks=true,      
    urlcolor=blue,
    citecolor=blue,
    linkcolor=blue
}

\begin{document}
\title{Quantum walks of interacting Mott insulator defects with three-body interactions}
\author{Suman Mondal and Tapan Mishra}
\affiliation{Department of Physics, Indian Institute of Technology, Guwahati-781039, India}

\date{\today}

\begin{abstract}
Quantum walks of interacting particles may display non-trivial features due to the interplay between the statistical nature and the 
many-body interactions associated to them. We analyze the quantum walk of interacting defects on top of an uniform bosonic Mott insulator at  
unit filling in an one dimensional graph. 
While the quantum walk of single particle defect shows trivial features, the case of two particles exhibits interesting phenomenon of quantum walk reversal
as a function of additional onsite three-body attractive interactions. In the absence of the three-body interaction a quantum walk of 
pairs of particles is obtained and 
as the strength of the three-body interaction becomes more and more attractive, the independent particle behavior in quantum walk appears. 
Interestingly, further increase in the three-body interaction leads to the re-appearance of the quantum walk associated to a pair of particles.
This quantum-walk reversal phenomenon is studied using the real-space density evolution, 
Bloch oscillation as well as two-particle correlation functions.

\end{abstract}





\maketitle
\section{Introduction} Dynamical systems often exhibit exotic physical phenomena due to the quantum nature of the particles. Understanding such 
phenomena in complex systems has been a topic of great interest both from theoretical and experimental point of view. Quantum walk(QW) is one of such 
phenomena which is the quantum analogue of the classical random walk has attracted enormous attention in recent years due to its relevance to 
physical and biophysical applications~\cite{Kempe2003}. The underlying physics arising from the wave-function overlap enables quantum mechanical particle to 
access various paths to optimize the motion on a graph showing a linear propagation of correlation limited by the Lieb-Robinson bound~\cite{Lieb-Robinson2006}.
This very idea of optimization is considered to be the key to develop efficient quantum 
algorithms. In the last decade, the quantum walks have been observed in various systems such as trapped ions, neutral atoms, 
photons in photonic lattices 
and waveguides, biological systems etc~\cite{Schmitz2009,Zahringer2010,Karski2009,Weitenberg2011,Fukuhara2013,Manouchehri2014,Hoyer2010,Mohseni2008} in 
the single particle level. 
Considerable efforts have been made to understand the 
effect of interactions on the QW of more than one indistinguishable particles in various systems such as quantum gases in optical lattice ~\cite{Greiner_walk},
correlated photon pairs~\cite{Peruzzo2010,Lahini_walk} and superconducting qubits~\cite{Ye2019,Yan2019}. In the interacting systems, the combined effect 
of quantum correlation and interaction may yield novel scenarios in the phenomenon of quantum walks as a result of the Hanburry-Brown and Twiss(HBT)
interference and 
Bloch oscillation~\cite{Bromberg2009,Peruzzo2010,Sansoni2012,Broome2013,Spring2013,Tillmann2013,Crespi2013,Greiner_walk,Zakrzewski2017,blochoscillation}.

In recent years, remarkable progress has been made in the experimental front in various systems to understand the quantum many-body effects of 
interacting particles. The ease of controlling the system parameters has paved the path to understand several 
complex phenomena in nature. One of such systems is the famous Bose-Hubbard model which deals with the dynamics of 
bosons in periodic potentials~\cite{Fisher1989}. Despite it's simplicity,
it has been shown to exhibit various fundamental properties such as the famous phase transition from a 
superfluid(SF) phase where the bosons are 
completely delocalized over the entire lattice to a localized Mott insulator(MI) phase~\cite{jaksch}. 
The experimental observation of this SF-MI transition~\cite{GreinerBloch}
in optical lattices with ultracold bosons has opened up a new avenue to explore numerous novel scenarios 
based on different variants of the Bose-Hubbard(BH) model in terms of higher
order local interactions~\cite{Will2010,Mark2011,Ueda_2011,Lompe2010,Tscherbul_2011,Sansone2012,Daley2009,Petrov2,Petrov1,TSowinski1,TSowinski2,TSowinski3,TSowinski4}, 
long range interactions~\cite{Boninsegni2012,Wolfgang2017,Léonard2017,Ferlaino2019}, 
artificial magnetism~\cite{Dhar2012,Dhar2013,Aidelsburger2013,Oktel2007,Giamarchi2001,Yasunaga2007,Petrescu2013,Sebastian2015}, cavity QED
~\cite{Hartmann2010,Hartmann2016,LeHur2016,Houck2012}, non-equilibrium phenomena~\cite{Mukund2011} etc. 
\begin{figure}[t]
\begin{center}
\includegraphics[width=.95\columnwidth]{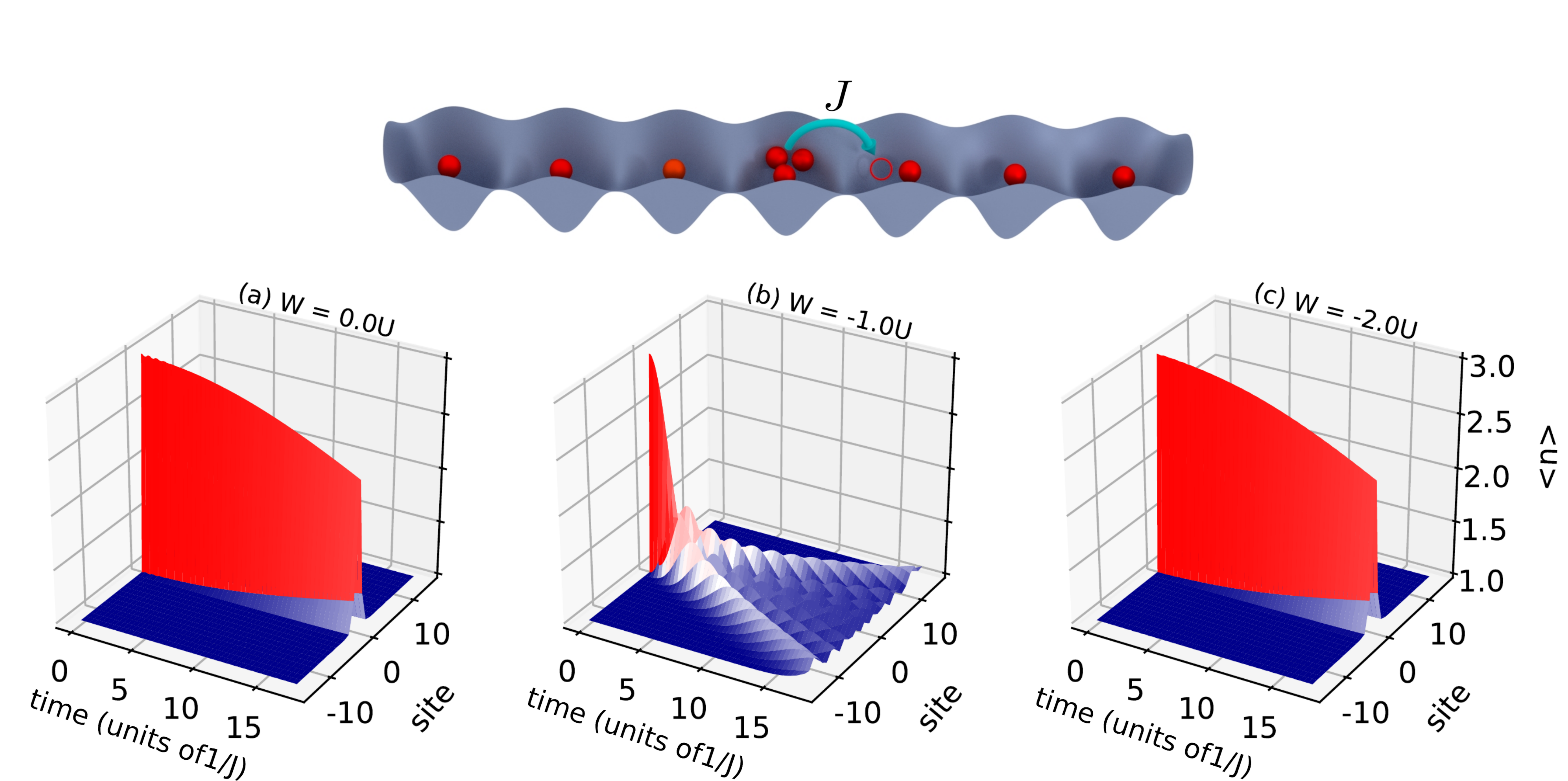}
 \end{center}
\caption{(Color online)Top panel: Figure shows the initial state which is two particles on top of a bosonic 
Mott insulator in one dimension at unit filling i.e. $a_0^{{\dagger}^2}|$MI$_1\rangle$. Here $J$ denotes the hopping strength.
Bottom panel: Phenomenon of QW reversal shown as a function of the three-body attractive interaction. }
\label{fig:lattice}
\end{figure}

Recently, the BH model has been 
analysed  to understand the quantum random walk of interacting particles in different physical 
systems~\cite{Lahini_walk,Greiner_walk,Yan2019,Zakrzewski2017,Childs2013}. 
In most of the cases, the focus was to study the QW with an initial state of particles in empty lattice driven 
by the competing two-body interactions. However, the effect of higher order local interaction may have signifiant impact on the 
QW of bosonic systems. On the other hand a natural question can be asked about the effect of the interaction coming from a lattice with 
occupied particles instead of empty sites. In this context we
investigate the continuous time QW of two interacting defects on top of an MI phase in one dimensioal BH model. 
We consider a different type of initial state where two defects are located initially at the 
same site on top of an perfect one dimensional Mott insulator at unit filling as shown in the top panel of Fig.~\ref{fig:lattice}. 
Motivated by the   
recent experimental progress in various systems such as cold atoms and circuit QED setups, we try to 
uncover the physics due to the enhancement of quantum effects in an interacting multi-particle system in 
the context of the BH model with local two and three-body interactions. 
Before going to the details about the results in the following sections, we briefly mention our important findings here.
We show that for fixed values of the two-body repulsion, a gradual increase in the attractive three body interaction 
results in the phenomenon of 
QW reversal as depicted in Fig. \ref{fig:lattice}(a-c). To be more specific, we find that initially the defects pair up and perform QW 
when the three-body interaction is vanishingly 
small. As the attractive three-body interaction increases, the pair tends to dissociate into mobile defects and the system 
exhibits independent particle QW. 
Further increase in the three-body attraction, results in an interesting phenomenon of QW reversal of the defect pairs, 
which will be discussed in detail in the following. 
%

The rest of the paper is organised in the following way. In section \ref{sec:mm}, we discuss about the model considered and 
method employed in this study. In Sec. \ref{sec:re}, the results are discussed in great detail and finally we conclude in Sec.\ref{sec:co} .

\section{Model and method}
\label{sec:mm}
The model which describes this system under consideration is the modified Bose-Hubbard model which is given by ;
\begin{eqnarray}
\mathcal{H}=&-&J\sum_{\langle i,j\rangle}(a_{i}^{\dagger} a_{j}+H.c.)+{{U}\over{2}}\sum_{i}n_i(n_i-1)\nonumber\\
&+&\frac{W}{6}\sum_i n_i(n_i-1)(n_i-2)
\label{eq:ham}
\end{eqnarray} 
where $a_{i}^{\dagger}$($a_{i}$) is the creation(annihilation) operator 
and $n_i=a_{i}^{\dagger} a_{i}$ is the number operator at $i$'th site. Here, $J$ is the hopping matrix
element and $U(W)$ is the two(three)-body onsite interaction energy. 
In the following, we discuss the QW of the MI defects in the presence of attractive three-body interaction $W$.

This scenario considered here is completely different from the quantum-walk of interacting bosons already discussed in the 
literature\cite{Greiner_walk,Lahini_walk,Zakrzewski2017}.
The very difference is that the quantum walkers interact with 
themselves as well as with the background bosons of the MI state. Although the interactions experienced from the 
background bosons in the MI state are uniform throughout the lattice, we will show that 
this background plays an important role in 
revealing interesting physics.

\begin{figure}[b]
\begin{center}
\includegraphics[width = 1\linewidth]{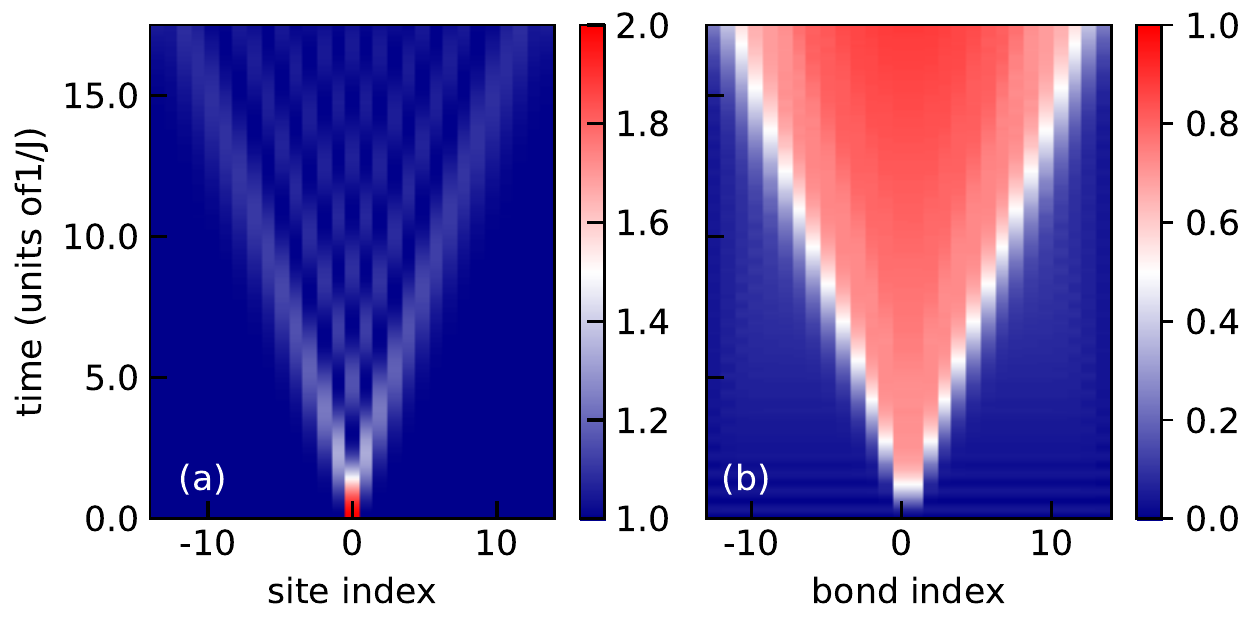}
\caption{(Color online)(a)Time evolution of $\langle n_i \rangle$ of a single defect with $J=0.2$ and $U = 10$ on an MI 
background of length $L = 29$ sites. (b) Propagation of
entanglement entropy $S$ shows the linear spread of information.}
\label{fig:qwsingle}
\end{center}
\end{figure}

\begin{figure*}[tb]
  \includegraphics[width=\textwidth]{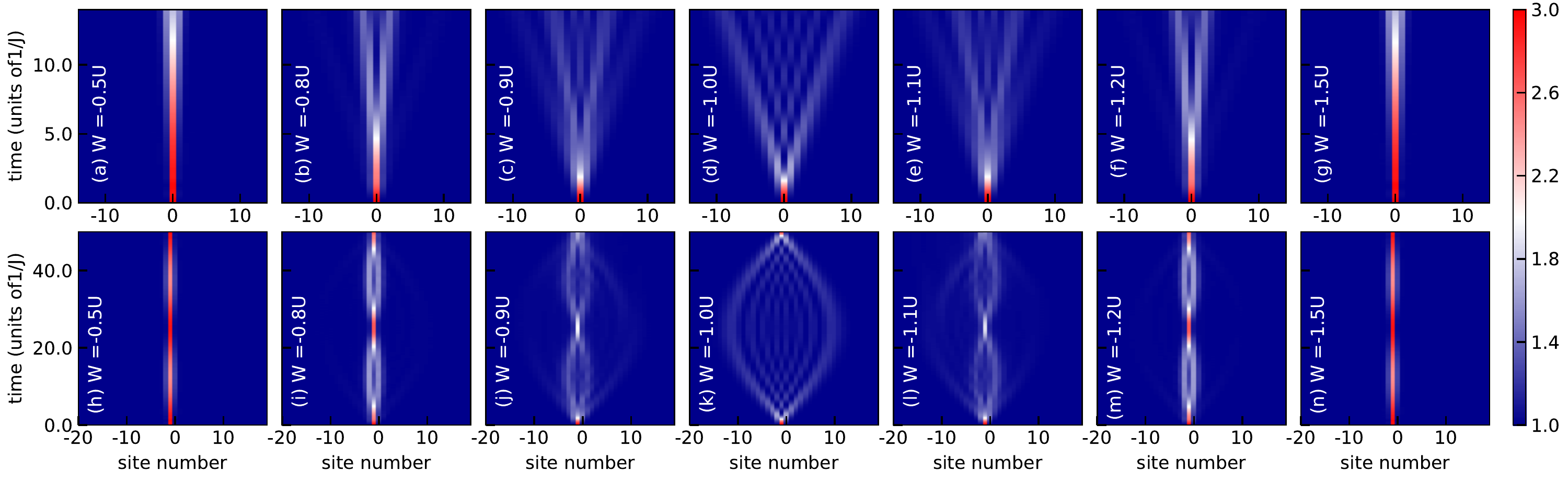}
  \caption{Figure shows the phenomenon of QW reversal of the pair of defects shown in terms of the density 
  evolution on a homogeneous(a-g) and tilted(h-n) lattice. In the case of tilted lattice the period doubling in 
  the Bloch oscillation clearly indicates the QW of bound defect pair.}
  \label{fig:qwpair}
\end{figure*}

Our starting point is a perfect MI state at unit filling with two defects created at the central 
site in an one dimensional periodic potential i.e. 
\begin{equation}\label{eq:ini1}
|\Psi_{0}\rangle=(a_0^\dagger)^2|MI_1 \rangle=|.~.~.~1~1~3~1~1~.~.~.\rangle
\end{equation}
The MI state is a result of large onsite repulsion $U$ compared to the hopping amplitude $J$. 
Note that in the process of quantum-walk there is a ballistic 
expansion of the particle wave function i.e. the probability of finding the particle at a specific distance from the starting point grows proportional to 
the diffusion time $t$. In contrast, for the classical case the probability grows diffusively as $\sqrt{t}$. Our focus is to understand the signatures 
of the QW from the particle density propagation which is defined as the expectation value 
\begin{equation}
 n_i(t)=\langle a_i^{\dagger}a_i\rangle
 \label{eq:ni}
\end{equation}
with the 
time evolved state $|\Psi(t)\rangle$ and the two particle correlation function 
\begin{equation}
 \Gamma_{ij} = \langle a_i^\dagger a_j^\dagger a_j a_i\rangle
 \label{eq:gamma}
\end{equation}
at fixed time which are accessible in 
recent experiments. Due to the large number of particles involved in the system, exact 
solution of the Schr\"{o}dinger equation with the Bose-Hubbard model is difficult. Hence, the dynamical evolution of the initial state is done by using the Time-Evolving 
Block Decimation(TEBD) method using the Matrix Product States(MPS)~\cite{OSMPS2} with maximum bond-dimension of $500$. This method is well suited for 
one-dimensional systems with local interactions~\cite{OSMPS1}. In our calculation we scale all the physical quantities by setting $J=0.2$. 

\section{Results}
\label{sec:re}
Before addressing the QW of a pair of defects we will show the QW of a single defect for completeness. The initial state in 
this case is \\
\begin{equation}
 |\Psi(0)\rangle=a_0^\dagger |MI_1\rangle=|.~.~.~1~1~2~1~1~.~.~.\rangle
 \label{eq:ini2}
\end{equation}
A single particle on top of an MI background will 
experience an uniform interaction and hence the system is identical to the QW of single particle in an empty lattice. Hence, one expect a typical ballistic 
expansion of $\langle n_i\rangle$ over the time $t$ during the process of evolution as shown in Fig.~\ref{fig:qwsingle}(a). 
We also compute the propagation of single defect entanglement entropy at $i$-th bond $S_i=$-Tr$(\rho_i $log$\rho_i)$ which shows a light-cone like spread of the information as 
depicted in Fig.~\ref{fig:qwsingle}(b). Here $\rho_i$ is the reduced density matrix defined at $i$-th bond connecting two part of the system.

\subsection{Density evolution}
\begin{figure*}[!t]
\begin{center}
\includegraphics[width = 1\linewidth]{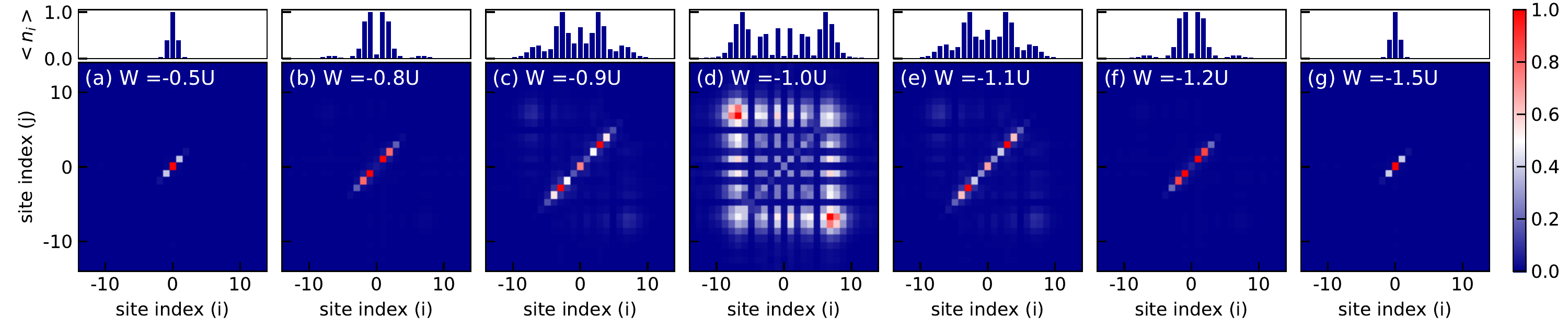}
\caption{(Color online)Figure shows normalized two particle correlation functions $\Gamma_{ij}$ plotted with respect to the positions $i$ and $j$ after
a time evolution of $11.25$ in units of ($1/J$) for $t=0.2$, $U = 10$ for different values of $W/U
=-0.5,~-0.8,~-0.9,~-1.0,~-1.1,~-1.2,~-1.5$(a-g) on a
homogeneous one dimensional lattice of length $L = 29$. Density distribution is shown on top of each correlation plot.}
\label{fig:corr}
\end{center}
\end{figure*}
In this section we discuss the QW of two interacting defects by analysing the real space density evolution 
using Eq.~\ref{eq:ni}. As mentioned before we start from the initial state with two 
defects on top of a perfect Mott insulator as shown in Eq.~\ref{eq:ini1}. For our analysis we consider $J=0.2$ and $U=10$ which makes the ratio $U/J=50$. 
With this 
ratio the two-particle repulsion between the bosons are very strong. We study the QW of such a system by systematically 
varying the three-body attraction
$W$ from very small to a very large value compared to $U$ which is shown in Fig.~\ref{fig:qwpair}(a-g). It can be seen that for $W = -0.5U$, 
the density evolution shows a slow propagation of the quantum walker although it is ballistic in nature. This indicates the 
QW of a slow moving particle in contrast to the independent particle QW. At this point 
, by increasing the three-body attraction, the density distribution gradually 
spreads and moves towards the boundaries of the lattice at a faster rate. At some intermediate values of $W$ ($W < -U <  W$), 
two different cones appear in the QW. In this regime of interaction, the signatures of both slow and fast moving particles are visible. 
Exactly at $W=-U$, the system exhibits a QW similar to that of the non-interacting particles(compare with Fig.~\ref{fig:qwsingle}). 
Interestingly, further increase in the three-body attraction after $W=-U$, 
slowly traces back through the intermideate phases appeared in $-W<U$ region to the original scenario (i.e. $W=-0.5U$), where we see the feature 
similar to the QW of a slow moving particles. In other words the change in $W$ in one direction introduces a QW reversal phenomena in the system.

\subsection{Bloch oscillation}
At this state it is difficult to ascertain about the different situations shown in Fig.~\ref{fig:qwpair}(a-g). So, in order to understand 
the nature of these two extreme situations we exploit the physics of the Bloch oscillation, which is the periodic breathing 
motion of particle in position space. This is an interesting manifestation of 
particle motion in a periodic potential subjected to an external force~\cite{blochoscillation}. This external force can be incorporated in model (1) 
as a constant tilt or gradient of the form 
\begin{equation}
 H_{tilt}=\lambda\sum_i i n_i
\end{equation}
Under the influence of this tilt potential, the particle undergoes a Bloch oscillation with period $\tau = 2\pi/\lambda$. 
We solve the model(1) with this additional term $H_{tilt}$ and study the density evolution 
for various values of $W$ as considered in Fig.~\ref{fig:qwpair}(a-g) with a tilt of {$\lambda=0.02 \times 2\pi$}. Interestingly, we see distinctly different features in the Bloch oscillations and 
a reversal phenomena as shown in Fig.~\ref{fig:qwpair}(h-n). It is interesting to note that for small and large values of $W$ the time period of oscillation 
are half that of the one at $W=-U$ (which corresponds to a independent particle type evolution). Note that the frequency doubling in this case 
is a typical signature of the Bloch oscillation of a pair of particles as already discussed 
in Ref.~\cite{Greiner_walk,Zakrzewski2017,DiasPRB2007,FlachPRA2010}. For intermediate values of $W$ ($W>-U$ and $W<-U$) 
there exists two types of oscillations with two 
different time periods. In this regime, it appears that both single and double occupancy state are energetically favorable. From this signature it is now easy to ascertain that 
the system exhibits a QW of a bound pair in the beginning when $W=-0.5U$ and gradually the pair tends to dissociate and the defects perform QW independently at $W \sim -U$. 
Counter intuitively, for larger values of $W$ the QW of pair reappears showing a reversal of QW phenomena as a function of $W$.

From the above analysis, it is evident that the quantum walker is a pair of defects for small and large values of $W$ compared to $U$. 
The pair which appears for $W=0$ can be thought of as a repulsively bound 
pair on top of the MI$_1$ state which is similar to the one observed in the quantum gas experiment by Winkler \textit{et. al.}~\cite{Winkler2006}. In this 
case, the MI$_1$ phase acts as a uniform background and hence the defect pair experiences uniform repulsions from all the sites. In such a case the 
velocity of the walker becomes extremely small as can be seen from the Fig.\ref{fig:qwpair} (upper panel). However, when the value of 
$W$ increases, the effective local interaction reduces gradually due to the attractive nature of $W$. Hence, the repulsively bound pair tends to dissociate 
into single particles and therefore, we see enhanced group velocity of propagation which corresponds to independent particle QW. However, the mechanism for the QW of pair in the large $W$ 
regime is completely different from the one for vanishing $W$. In this regime, the pairing of defects is due to the combined effect of the 
interactions $U$ and $W$. This is altogether a different kind of mechanism to establish repulsively bound pairs which can be understood as follows. 
When $W$ is very large 
and attractive compared to the other energy scales of the system then ideally the system prefers to form a trimer(three particle bound state). This trimer may consists of 
the two defect bosons and one from the MI background. However, because of the uniform repulsion from all the sites 
due to the MI state the two defects may rather prefer 
to move freely throughout the system as bound pair ~\cite{Manpreet_pair,sumanpair}. It is to be noted that although the pair 
formation mechanism for both the cases($W = 0$ and $\neq 0$) are 
different, the signatures in the quantum 
walk are identical in nature.

\subsection{Two particle correlations}
At this stage, to further substantiate the physics presented above we utilize the two-particle correlator $\Gamma_{i,j}$ defined in Eq.\ref{eq:gamma},
which sheds light on the quantum coherence of the two particles. 
It is well known that if two particles perform QW together, then HBT interference occurs which strongly depends 
on the statistical nature of the particles. However, in the present case, since the quantum walkers originate from the same site, 
the HBT interference are forbidden.
We compute $\Gamma_{i,j}$ after 
an evolution time of $t=11.25$(units of $1/J$) for different values of $W$ as shown in the bottom panel of Fig.~\ref{fig:corr}(a-g). 
We have considered a reduced basis to
define the two particle correlator $\Gamma_{ij}$ and number operator $n_i$ where we subtract the contributions from the MI$_1$
background. The mapping between the initial and the reduced on-site basis reads $\{|n\rangle\}
\rightarrow \{|n-1\rangle\}$ for $n>0$.
The correlation functions(particle densities) of each plot of Fig.\ref{fig:corr} are
normalized by their largest respective values so that each plot can
share the same scale from $0$ to $\Gamma_{max}$(or zero to one). One can clearly see that when the ratio $W/U$ is very 
small the diagonal weights of the correlation matrix are dominant indicating 
the QW of repulsively bound pair(Fig.~\ref{fig:corr}(a)). Increasing the value of $W/U$ to a very large limit recovers 
the similar behavior in the correlation matrix corresponding to a QW of bound pair. 
However, at intermediate regime of the ratio $W/U$, the off-diagonal weights of the correlation matrix start to increase 
and eventually showing the signature of independent particle QW 
as shown in Fig.~\ref{fig:corr}(d). These signatures in the two particle correlators for independent and pair particle quantum walks 
are similar to the one obtained in recent experiment for two interacting 
particle quantum walk in empty lattice~\cite{Greiner_walk}. In the top panel of Fig.~\ref{fig:corr} we plot the normalized densities 
$\langle n_i \rangle$ which shows features complementing the two particle correlation behavior.

\section{Conclusions} 
\label{sec:co}
We analyses the QW of two interacting defects on a perfect MI$_1$ state in the context of the Bose-Hubbard model with 
both two-body repulsive and three-body attractive interactions. By fixing the onsite two-body interaction at a finite value and 
varying the three-body interaction from zero to large value we predict the phenomenon of QW reversal. We show that the two defects on top of the MI phase pair up and perform QW for small and large values of $W$. At intermediate strength of $W$, the defects behave like independent walkers in the QW. We rigorously discuss this process in the time evolution of real-space 
density distribution, Bloch oscillation and also two particle correlation function. This results shows a spontaneous QW reversal process in 
Mott insulator defects. 

The above findings are based on a simple Bose-Hubbard model with two and three-body interactions and one of the 
immediate platform where one can think 
of observing this QW reversal phenomena is quantum gas experiment in optical lattices. The simultaneous existence of 
both two and three-body interactions 
has been observed in recent experiment in optical lattices~\cite{Will2010}. Several theoretical proposals 
have been made to engineer and tune 
the three-body interaction in optical lattices~\cite{Tiesinga2009,Petrov1,Daley2009,Sansone2012}. 
Moreover, recent observation of 
QW with single-site addressing in interacting ultracold atoms in optical lattices~\cite{Greiner_walk}
have broadened the scope by many-fold. In the optical lattice setups, it can be possible to create an initial state proposed in this 
work by creating a Mott insulator phase at $n=3$ and selectively removing a pair of particles from every site except 
the central one. With the proposed mechanism to tune the two and three-body interactions in optical lattice, the time evolution of such 
initial state may reveal the quantum walk reversal phenomenon. On the other hand quantum simulations in circuit QED systems 
have attracted enormous attention in recent years due to the flexibility to design and control strong non-linearities and interactions with 
microwave radiation and artificial atoms. Very recently, strongly correlated quantum walks with a 12-qubit superconducting circuit 
has been observed in experiment~\cite{Yan2019}. In practice two-level artificial atoms are considered in the quantum simulations with 
circuit QED setups. However, a recent experimental proposal shows that it is possible to control the two and three-body interactions by considering a 
fluxonium qubit~\cite{Hafezi2014} where the first and second excitation levels are of equal energy and the third one can be controlled by detuning it from the 
first two. This scenario results in a two- and three-body interacting Bose-Hubbard model. In such a scenario the above predicted physics of QW 
reversal can be observed in the current state-of-the art experiments based on quantum gases in optical lattice or circuit QED systems. 
This result also opens up directions to study other interesting quantum mechanical phenomena such as the HBT 
interference effects~\cite{Bromberg2009,Peruzzo2010,Sansoni2012,Broome2013,Spring2013,Tillmann2013,Crespi2013,Lahini_walk,Greiner_walk} in 
such multi-body interacting quantum walks of defects.


\begin{acknowledgments}
We thank David Carpentier, Abhishek Dhar and Kanhaiya Pandey for useful discussions. We acknowledge useful suggestions from Daniel 
Jaschke and Arya Dhar related to the open source OSMPS software. 
The computational simulations were carried out using the Param-Ishan HPC facility at
Indian Institute of Technology - Guwahati, India.
T.M. acknowledges DST-SERB for the early career grant through Project No.
ECR/2017/001069.
\end{acknowledgments}

\bibliography{references}

\end{document}